\documentclass[prd,twocolumn,a4paper,superscriptaddress,floatfix]{revtex4}
\usepackage{graphicx}
\usepackage{amssymb}
\usepackage{amsmath}

\begin{document}
\newcommand{\be}{\begin{equation}}
\newcommand{\ee}{\end{equation}}
\newcommand{\bq}{\begin{eqnarray}}
\newcommand{\eq}{\end{eqnarray}}
\newcommand{\bsq}{\begin{subequations}}
\newcommand{\esq}{\end{subequations}}
\newcommand{\bc}{\begin{center}}
\newcommand{\ec}{\end{center}}
\newcommand\lapp{\mathrel{\rlap{\lower4pt\hbox{\hskip1pt$\sim$}} \raise1pt\hbox{$<$}}}
\newcommand\gapp{\mathrel{\rlap{\lower4pt\hbox{\hskip1pt$\sim$}} \raise1pt\hbox{$>$}}}
\newcommand{\dpar}[2]{\frac{\partial #1}{\partial #2}}
\newcommand{\sdp}[2]{\frac{\partial ^2 #1}{\partial #2 ^2}}
\newcommand{\dtot}[2]{\frac{d #1}{d #2}}
\newcommand{\sdt}[2]{\frac{d ^2 #1}{d #2 ^2}}
\newcommand{\vv}[0]{{\bar v}}
\newcommand{\cc}[0]{{\tilde c}}
\newcommand{\ave}[1]{\left< #1 \right>}
\newcommand{\Ogw}[0]{\Omega_{\rm gw}}
\newcommand{\lp}[0]{{\rm f}}

\title{Full analytical approximation to the stochastic gravitational wave background generated by cosmic string networks}

\author{L. Sousa}
\email[Electronic address: ]{Lara.Sousa@astro.up.pt}
\affiliation{Instituto de Astrof\'{\i}sica e Ci\^encias do Espa{\c c}o, Universidade do Porto, CAUP, Rua das Estrelas, PT4150-762 Porto, Portugal}
\affiliation{Centro de Astrof\'{\i}sica da Universidade do Porto, Rua das Estrelas, PT4150-762 Porto, Portugal}

\author{P. P. Avelino}
\email[Electronic address: ]{pedro.avelino@astro.up.pt}
\affiliation{Departamento de F\'{\i}sica e Astronomia, Faculdade de Ci\^encias, Universidade do Porto, Rua do Campo Alegre 687, PT4169-007 Porto, Portugal}
\affiliation{Instituto de Astrof\'{\i}sica e Ci\^encias do Espa{\c c}o, Universidade do Porto, CAUP, Rua das Estrelas, PT4150-762 Porto, Portugal}
\affiliation{Centro de Astrof\'{\i}sica da Universidade do Porto, Rua das Estrelas, PT4150-762 Porto, Portugal}
\affiliation{School of Physics and Astronomy, University of Birmingham,Birmingham, B15 2TT, United Kingdom}

\author{G. S. F. Guedes}
\email[Electronic address: ]{gguedes@lip.pt}
\affiliation{Laborat\'{\o}rio de Instrumenta{\c c}\~{a}o e F\'{\i}sica Experimental de Part\'{\i}culas, Departamento de F\'{\i}sica da Universidade do Minho, Campus de Gualtar, 4710-057 Braga, Portugal}

\begin{abstract}
We derive a full analytical approximation to the stochastic gravitational wave background generated by the loops that are produced throughout the cosmological evolution of cosmic string networks. We show that this approximation not only predicts the amplitude of the radiation-era plateau exactly, but also provides a good fit to the high-frequency cut-off and to the low-frequency peak generated by the loops that decay during the matter era, irrespective of cosmic string tension and of the length of loops created. We then find that it provides a good quantitative description of the full stochastic gravitational wave background across the relevant frequency range.

\end{abstract} 
\maketitle

\section{Introduction}

The production of cosmic string networks as a consequence of symmetry-breaking phase transitions is a crucial prediction of several beyond the standard model theories, including grand unified scenarios~\cite{Kibble:1976sj}. These networks are generally expected to survive throughout cosmological history, potentially leaving behing characteristic observational signatures which can be used to probe the underlying particle physics (for a review see~\cite{Hindmarsh:1994re,Vilenkin:2000jqa}). One such signature is a characteristic Stochastic Gravitational Wave Background (SGWB) generated by cosmic string loops~\cite{Vilenkin:1981bx,Hogan:1984is,Accetta:1988bg}. These loops are generally expected to be copiously produced throughout cosmological history as a result of the frequent interactions between strings. After creation, however, they detach from the network and decay by emitting their energy in the form of Gravitational Waves (GWs). It is the superposition of the individual transient signals of these cosmic string loops that gives rise to this background of gravitational radiation.

The recently inaugurated era of GW Astronomy~\cite{Abbott:2016nmj,TheLIGOScientific:2017qsa} has opened new possibilities to probe the SGWB generated by cosmic string loops. The Laser Space Interferometer Space Antenna (LISA), in particular, may be able to detect cosmic string networks up to tensions of $G\mu=\mathcal{O}(10^{-17})$~\cite{Auclair:2019wcv} (where $G$ is the universal gravitational constant), an improvement of about six orders of magnitude over current constraints~\cite{Lentati:2015qwp}. LISA will then be an ideal instrument to probe this background and it is expected to either result in its detection or in a significant tightening of the constraints on cosmic string-forming scenarios. In any case, accurately characterizing the SGWB generated by cosmic string networks is pivotal to use the current and upcoming GW detectors to their full potential and, for this reason, this subject has been generating considerable interest~\cite{Sanidas:2012ee,Binetruy:2012ze,Kuroyanagi:2012wm,Sousa:2013aaa,Sousa:2016ggw,Blanco-Pillado:2017oxo,Ringeval:2017eww,Cui:2018rwi,Jenkins:2018lvb,Auclair:2019jip,Gouttenoire:2019kij}.

The characterization of the SGWB produced by cosmic strings often relies on numerical tools. However, multiple computations of the spectrum, covering a wide parameter space, are generally necessary to confront different cosmic string scenarios with observational data. As a result, this numerical approach may be, in some instances, rather slow and computationally costly. In this paper, we derive an analytical approximation to the SGWB generated by cosmic string networks that accurately quantifies the full spectrum. It rigorously describes the main features of the SGWB spectrum produced by a network of cosmic strings such as the low- and high-frequency cut-offs, the peak originated by loops that decay in a matter-dominated era and the plateau resulting from loops that decay in a radiation era. This analytical approximation is accurate independently of the size of loops and for different values of the string tension, thus allowing for the quick and efficient calculation of the SGWB for a wide range of parameters.

This paper is organized as follows. In Sec.~\ref{sec-strings}, we introduce the Velocity-dependent One Scale (VOS) model to describe the cosmological evolution of cosmic string networks. The emission of GWs by cosmic string loops is briefly reviewed in Sec.~\ref{sec-SGWB}, and we also introduce the method for computation of the SGWB spectrum therein. In Sec.~\ref{sec-general}, we derive the general shape of the spectrum and compute the amplitude of the radiation-era plateau of the spectrum. In Sec.~\ref{sec-ldf}, we provide an analytical description for the loop distribution function for a network in which all the loops created at an given time have the same length and show that this model may be used to construct the loop distribution function of more complex models in which loop production happens at more than one lengthscale. In Sec.~\ref{sec-analytic}, we provide a detailed derivation of an analytical approximation to the SGWB generated by realistic cosmic string networks and we compare this approximation to spectra obtained numerically in Sec.~\ref{sec-comp}. We then conclude in Sec.~\ref{sec-conclusions}.
Throughout this paper we will use natural units with $c=\hbar=k_B=1$ (unless explicitly stated otherwise).

\section{The cosmological evolution of cosmic string networks\label{sec-strings}}

The cosmological evolution of cosmic string networks may be described statistically, on sufficiently large scales, by following the evolution of two variables: the characteristic length $L$ --- which is a measure of the energy density of the network $\rho=\mu/L^2$ --- and the Root-Mean-Squared (RMS) velocity $\vv$~\cite{Martins:1996jp,Martins:2000cs}:

\bq
\dtot{\vv}{t} & = & \left(1-\vv^2\right)\left[\frac{k(\vv)}{L}-2H\vv\right]\,,\label{eqvosv}\\
\dtot{L}{t} & = & \left(1+\vv^2\right)HL+\frac{\cc}{2}\vv\label{eqvosL}\,,
\eq
where
\be
k(\vv)=\frac{2\sqrt{2}}{\pi}\left(1-\vv^2\right)\left(1+2\sqrt{2}\vv^2\right)\frac{1-8\vv^6}{1+8\vv^6}\label{curv}
\ee
is a phenomenological parameter that accounts, to some extent, for the effects of small-scale structure on the long strings~\cite{Martins:2000cs}, $\cc=0.23\pm 0.04$~\cite{Martins:2000cs} is a parameter quantifying the efficiency of the energy-loss mechanism of the network, $H=(da/dt)/a$ is the Hubble parameter, and $a$ is the cosmological scale factor. Eqs.~(\ref{eqvosv}) and~(\ref{eqvosL}), known as the Velocity-dependent One-Scale (VOS) model, provide a thermodynamical description of the cosmological evolution of cosmic string networks. (Note however that here we have not included the effects of the frictional forces caused by the scattering of the particles of the background plasma by cosmic strings, which are included in the original VOS model. These frictional forces are only expected to be relevant at early cosmological times before the production of gravitational waves becomes significant and hence they may, in general, be neglected in this context.)

The interactions between strings play a crucial role in cosmic string network dynamics. When two strings collide, they generally exchange partners and reconnect. This process, known as intercommutation, not only leads to the formation of small-scale structure on the string but also often results --- when a string self-intersects or two strings intersect simultaneously in two points --- in the formation of closed cosmic string loops. After formation, these loops detach from the cosmic string network and decay radiatively. Cosmic string interactions then result in a stream of energy loss by the cosmic string network that may be written as~\cite{Kibble:1984hp}:

\be
\left.\frac{d\rho}{dt}\right|_{\rm loops}=\cc\vv\frac{\rho}{L}\label{energyloss}\,.
\ee

As a result of this energy loss, the cosmic string network evolves towards a linear scaling regime during which its characteristic length grows proportionally with physical time and its energy density remains a fixed fraction of the energy density of the cosmological background. In fact, one may see that a regime of the form

\be
L=\xi t\quad\mbox{and}\quad \dtot{\vv}{t}=0\label{sca}\,,
\ee
with
\be
\xi^2=\frac{k\left(k+\cc\right)}{4\beta\left(1-\beta\right)}\quad\mbox{and}\quad \vv^2=\frac{k}{k+\cc}\frac{1-\beta}{\beta}\label{sca-par}
\ee
is an attractor solution of the VOS equations for $a\propto t^\beta$ and $0<\beta<1$.

\section{The stochastic gravitational wave background generated by cosmic string networks\label{sec-SGWB}}

When cosmic string loops detach from the long string network, they start to oscillate relativistically under the effect of their tension and decay by emitting gravitational radiation. The frequencies of the GWs emitted by a loop are given by~\cite{Burden:1985md,Allen:1991bk}:

\be 
f_j=\frac{2j}{l}\,,
\ee
where $l$ is the length of the loop at the time of emission and $f_j$ is the frequency corresponding to the $j$-th harmonic mode. The distribution of power in the different harmonics is determined by the small-scale structure of the loops. In fact, it was shown~\cite{Vachaspati:1984gt,Binetruy:2009vt} that, in the large $j$ limit, the power emitted in each mode scales as $j^{-4/3}$ if the loop contains points where the velocity is locally $1$ known as cusps; as $j^{-5/3}$, if it has kinks (which are discontinuities in the tangential vector introduced by the intercommutation process); and $j^{-2}$ when kink-kink collisions occur. Note however that, at any given time, there are several loops with different shapes and structure and one may, in general, assume that, on average, the power emitted in each mode is

\be
\dtot{E_j}{t}=P_j G\mu^2\,, 
\ee
where

\be 
P_j=\frac{\Gamma}{\mathcal{E}}j^{-q}\label{powerspec}\,,
\ee
is the averaged loop power spectrum, $\mathcal{E}=\sum_m ^{n_*} m^{-q}$, $q=4/3\,,5/3\,,2$ for loops with cusps, kinks or for kink-kink collisions respectively, and $\Gamma\sim 50$ is a parameter quantifying the efficiency of GW emission~\cite{Vilenkin:1981bx,Scherrer:1990pj,Quashnock:1990wv,Blanco-Pillado:2017oxo}. In principle, when computing the SGWB generated by cosmic string loops, one should consider the contribution of all the harmonic modes of emission (i. e., $n_*=\infty$). Note however that it was shown in \cite{Sanidas:2012ee} that, in general, one only needs to consider modes up to a cut-off $n_*=10^3\,,10^5$ for $q=4/3\,,2$ respectively, since the inclusion of higher harmonics has a negligible impact on the shape of the spectrum.

As a result of GW emission, cosmic string loops lose energy, on average, at a roughly constant rate

\be
\dtot{E}{t}=\Gamma G\mu^2\,,
\ee
where $E=\mu l$ is the loop energy, and consequently their length decreases until they eventually evaporate. Loops then give rise to a transient signal of GWs. However, at any given time of the evolution of the network, there is a large number of loops emitting GW bursts in many different directions. The superposition of all these emissions is expected to give rise to a SGWB~\cite{Vilenkin:1981bx,Hogan:1984is,Brandenberger:1986xn,Accetta:1988bg}.

The amplitude of the SGWB generated by cosmic string loops is generally characterized through the spectral energy density of GW (in units of critical density $\rho_c=3H_0^2/8\pi G$):

\be 
\Ogw(f,q,n_*) =\frac{1}{\rho_c}\dtot{\rm{\rho}_{gw}}{\log f}\,.
\label{spectrum}
\ee

If we separate the contributions from the different harmonic modes, we may write

\be
\Ogw(f,q,n_*)=\sum_j^{n_*}\frac{j^{-q}}{\mathcal{E}}\Ogw^j(f)
\label{higher1}\,, 
\ee
where

\be 
\Ogw^j(f)=\frac{16\pi}{3}\left(\frac{G\mu}{H_0}\right)^2\frac{\Gamma}{f}\int_{t_i}^{t_0} j\, n\left(l_j\left(t'\right),t'\right)\left(\frac{a(t')}{a_0}\right)^5 dt'\label{spec}
\ee
is the contribution of the $j$-th harmonic mode of emission to the SGWB~\cite{Sanidas:2012ee,Sousa:2014gka}. Here, $t_i$ is the instant of time in which loop production becomes significant, which is often assumed to be at the end of the friction-dominated regime $t_i\sim t_{\rm pl}/(G\mu)^2$~\cite{Vilenkin:2000jqa} (where $t_{\rm pl}$ is the Planck time), and the subscript `0' is used to refer to the value of the corresponding parameter at the present time. The loop distribution function $n\left(l_,t\right) dl$ gives us the number density of string loops with lengths between $l$ and $l+dl$ that exist at the time $t$, and $l_j(t')=2ja(t')/fa_0$ is the physical length of the loops that radiate in the $j$-th harmonic mode at a time $t'$ GWs that have a frequency $f$ at the present time.

Note that, since one has

\be 
\Ogw^j(jf)=\Ogw^1(f)\,,\label{higher2}
\ee
one may straightforwardly construct $\Ogw^j(f)$ for an arbitrary harmonic mode $j$ once the contribution of the fundamental mode, with $j=1$, is computed. We will, therefore, restrict ourselves to the fundamental mode for (most of) the remainder of this article (and drop the superscript `1'). We will discuss the contribution of higher order modes in Sec.~\ref{sec-comp}.

\section{General considerations about the cosmic string SGWB\label{sec-general}}

Eq.~(\ref{energyloss}) implies that in a linear scaling regime, the decrease of the energy density of the network in an infinitesimal logarithmic time interval $d \log t$ due to loop production is proportional to $\rho \, d \log t$, with the average energy density $\rho$ of the cosmic string network scaling proportionally to $t^{-2}$. If the normalized distribution of loop lengths at the moment of creation scales linearly with the cosmic time and $\Gamma$ is a constant, the variation of the gravitational wave background energy density associated with loop emission in the time interval $[t_e,t_e+dt_e]$ is given by
\be 
d {\rho_{\rm gw[e]}} \propto \rho_e \, d \log t_e \propto t_e^{-2} d\log t_e\propto a_e^{-2/\beta} d \log a_e\,,
\ee
where the subscript `$e$' indicates that the quantities are evaluated at the time $t_e$, and the normalized probability density distribution ${\mathcal P}(f)$ of the emitted gravitational wave energy over frequency is independent of the emission time $t_e$ once the frequencies are rescaled in proportion to the inverse of $t_e$. The corresponding quantity at the present time (after including the volume expansion and redshift effects) is
\be
d {\rho_{\rm gw }} = d {\rho_{\rm gw[e]}} a_e^{4} \propto a_e^{(4\beta -2)/\beta} d \log a_e\,.
\ee

Let us consider the frequencies $f_e \propto 1/t_e$ corresponding to an arbitrary constant value of ${\mathcal P}$. An observed frequency at a time $t$ is related to the corresponding emitted frequency at the time $t_e$ through $f=a_e f_e \propto a_e/t_e \propto a_e^{(\beta-1)/\beta}$. Hence, $a_e \propto f^{\beta/(\beta-1)}$ with $d \log a_e \propto d \log f$, thus implying that
\be
\frac{d\rho_{\rm gw}}{d \log f} \propto f^{(4\beta -2)/(\beta-1)} \propto f^{2 (w-1/3)/(w+1/3)}  \,.
\label{slope}
\ee
Here, we have used the relation $\beta=2/(3(1+w))$, where $w$ is a constant equation-of-state parameter (equal to the ratio between the proper pressure and the proper density of the universe). The  result given in Eq. (\ref{slope}) is only valid for $w \ge 1/9$. For $w \le 1/9$ the spectrum  would be proportional to $f^{-1}$ --- the characteristic power spectrum of the gravitational radiation emitted at the end stages of the lifetime of the loops created throughout the cosmological history \cite{Guedes:2018afo} --- and is dominated by late-time contributions even at high frequencies. In summary, the  result given in Eq.  (\ref{slope}) (with the $f^{-1}$ cut-off for $w \le 1/9$) is independent of ${\mathcal P}(f_e)$ and, therefore, this is a simpler way of determining the general shape of the power spectrum for power law cosmologies considered in \cite{Cui:2017ufi,Cui:2018rwi}. 

Eq. (\ref{slope}) implies that for $w=1/3$ the power spectrum is flat, thus explaining the radiation-era plateau of the SGWB. A corollary of this result is that the normalization of the radiation-era plateau is independent of the shape of  ${\mathcal P}(f)$ --- only the total energy converted into gravitational waves matters, irrespective of its distribution over frequency. One may, therefore, take advantage of this property to compute the normalization of this plateau in a simple way.

Let us start by considering a loop created at an instant $t_b$ with a length $l_b=l(t_b)$ and assume the network is in a linear scaling regime in a background with $a\propto t^\beta$. The energy that arrives in form of GWs to an observer at $t_0$ is given by

\be
E_0=\int_{t_b}^{t_d}\frac{dE}{dt}\frac{a(t')}{a_0}dt'=E_b\frac{a_b}{a_0}\frac{\left(1+\epsilon\right)^{\beta+1}-1}{\epsilon\left(1+\beta\right)}\,,
\label{energy-loop0} 
\ee
where the subscript `b' is used to refer to the value of the corresponding quantity at the time of birth of the loop $t_b$, $E_b=\mu l_b$, and $t_d=(1+\epsilon)t_b$ is the time at which the loop completely evaporates. Here, we have also defined $\epsilon=l_b/(\Gamma G\mu t_b)$, which is independent of $t_b$ since we have assumed that the length of the loops at the time of creation scales, if the network is in a linear scaling regime, proportionally to cosmic time.

The spectral energy density of GWs emitted by the loops created at a time $t_b$ that reaches an observer in the present is then given by
\be
\frac{d {\rho_{\rm gw}}}{dt df}=\left.\frac{d\rho}{dt}\right|_{\rm loops}\frac{\left(1+\epsilon\right)^{\beta+1}-1}{\epsilon\left(1+\beta\right)}\left(\frac{a(t_b)}{a_0}\right)^4\mathcal{P}(f)\,.
\label{spectral-density}
\ee

The only ingredient missing to compute the amplitude of the plateau is the probability density distribution of the GW energy over frequency $\mathcal{P}$. It is straightforward to show that the normalized distribution of the GW energy density over emitted frequency $f_e$ is given by

\be
\mathcal{P}(f_e)=\left|\frac{dE}{df_e}\right|\Theta(f_e-f_{min,e})=\frac{f_{min,e}}{f_e^2} \Theta(f_e-f_{min,e})\,,
\label{pdf}
\ee
where $f_{min,e}=2/l(t_b)$ is the minimum frequency emitted by the loops and $\Theta(x)$ is a Heaviside function. As predicted, this is independent of $t_e$ provided $f_{min,e}$ is re-scaled in proportion to the inverse of $t_e$. Note that, as a loop emits GWs, its length decreases and consequently the frequency of emission increases. As a result, a loop is expected to emit GWs for all frequencies $f\ge f_{min,e}$ (hence the inclusion of the Heaviside function). Small loops, which have $l_b\ll \Gamma G\mu t_b$, decay effectively immediately on the cosmological timescale and thus, as shown in~\cite{Sousa:2014gka}, it is reasonable to assume that the entirety of the energy of the loops is radiated instantaneously at $t_b$. As a result, the probability density distribution $\mathcal{P}(f)$ of the emitted GWs over observed frequency $f$ is also given by Eq.~(\ref{pdf}), but with $f_e\to f$ and $f_{min,e}\to f_{min}=2 a_b/(l(t_b)a_0)$~\cite{Sousa:2014gka}. In the case of large loops, with $l\ge\Gamma G\mu t_b$, --- in which the approximation that loops radiate their energy instantaneously at the time of formation no longer holds --- GWs are emitted in distinct instants of time (between $t_b$ and $t_d$) and for this reason are redshifted by different amounts. This gives rise to a more complex $\mathcal{P}(f)$, whose explicit dependence on $a_b$ and $f$ depends on $\beta$. However, since the normalization of the plateau is independent on the explicit form of this function (provided we take Eq.~(\ref{energy-loop0}) into account), here for simplicity we chose to use the $\mathcal{P}(f)$ for small loops (given by Eq.~(\ref{pdf}), with $f_e\to f$ and $f_{min,e}\to f_{min}$).

Since loops emit GWs for all $f\ge f_{min}(t_b)$, the first loops that will contribute to the SGWB at a given frequency $f$ will be those created at a time $t_{min}$ for which $f_{min} (t_{min})=f$. If one considers a radiation-dominated universe, with

\be
H^2=H_0^2\Omega_r\left(\frac{a_0}{a(t)}\right)^4\label{radera}\,,
\ee
this instant is determined by

\be
\frac{a_0}{a_{min}}=\frac{\epsilon\Gamma G\mu}{4H_0\Omega_r^{1/2}}f\,, 
\ee
where $\Omega_r$ is the radiation density parameter at the present time. By using Eqs.~(\ref{spectrum}),~(\ref{energy-loop0}) and~(\ref{spectral-density}) with $\beta=1/2$, we find that the amplitude of the radiation-era plateau of the cosmic string SGWB is given by

\bq
\Ogw^{\rm plateau}(f) & = & \frac{32\pi}{9H_0^2}\frac{A_r}{f}\int_{t_{min}}^{+\infty}\left(\frac{a(t_b)}{a_0}\right)^5\frac{\left(1+\epsilon\right)^{3/2}-1}{\Gamma\epsilon^2 t_b^4}dt_b\nonumber\\
& = & \frac{128}{9}\pi A_r \Omega_r\frac{G\mu}{\epsilon}\left[\left(1+\epsilon\right)^{3/2}-1\right]\,,
\label{plateau}
\eq
where we have defined $A_r=\cc v_r/(\sqrt{2}\xi_r^3)$ --- which includes a factor of $1/\sqrt{2}$ to account for the effects of the redshifting of the peculiar velocities of loops~\cite{Vilenkin:2000jqa} --- and where $v_r$ and $\xi_r$ are the scaling constants during the radiation era (given by Eq.~(\ref{sca-par}), with $\beta=1/2$). The result in Eq.~(\ref{plateau}) is identical to the result of the full computation presented in~\cite{Auclair:2019wcv} (see also~\cite{Vilenkin:2000jqa}).

As to the SGWB created during the matter era, we shall see later in this article that, although the simple estimation in Eq.~(\ref{slope}) is generally accurate, the picture is somewhat more complex.

\section{Loop distribution function\label{sec-ldf}}

As we have seen, the loop distribution function is the crucial ingredient to compute the SGWB generated by cosmic string networks. In order to construct this function one may either resort to numerical simulations of cosmic string networks~\cite{Lorenz:2010sm,Blanco-Pillado:2013qja,Blanco-Pillado:2019tbi} or use analytical modeling~\cite{Kibble:1984hp,Caldwell:1991jj,DePies:2007bm,Sanidas:2012ee,Sousa:2013aaa} (see~\cite{Auclair:2019wcv} for a recent review). Here, since we aim to derive an analytical approximation to the spectrum, we shall opt for the later and follow the approach in \cite{Sousa:2013aaa}.

When constructing the loop distribution function analytically, it is generally assumed that the energy lost by the network (given in Eq.~(\ref{energyloss})) goes into the formation of cosmic string loops, so that

\be
\cc\vv\frac{\rho}{L}=\mu\int_0^\infty l \,\lp (l,t)dt\label{energybalance}\,, 
\ee
where the loop production function $\lp (l,t)$ represents the number density of loops with length between $l$ and $l+dl$ produced per unit time. As discussed in~\cite{Blanco-Pillado:2019vcs}, this assumption is crucial to get the correct normalization of the loop production function. Moreover, in this context, it is also often assumed that all of the loops have the same length at the moment of creation. Here, we shall assume that this length is determined by the characteristic length of the long string network at the moment of creation $t_b$~\cite{Sousa:2013aaa}:

\be
l(t_b)=\alpha L(t_b)\,, 
\ee
where $\alpha<1$ is a constant loop-size parameter which is generally treated as a free parameter. In reality, one does not expect all loops to be created with exactly the same size, but instead to follow a distribution of lengths. We will discuss the effects of relaxing this assumption later in this section.

Under this assumption, and taking into account the energy balance in Eq.~(\ref{energybalance}), one should have

\be
\lp (l,t)=\frac{\cc}{\sqrt{2} l}\frac{\vv(t)}{L(t)^3}\delta\left(l-\alpha L\right)\,, 
\ee
where we have also included the $1/\sqrt{2}$ correction factor that accounts for the energy lost due to the redshifting of the peculiar velocities of cosmic string loops~\cite{Vilenkin:2000jqa}.

Since, after formation, the size of loops decreases due to the emission of gravitational radiation roughly as

\be
l(t)=\alpha L(t_b)-\Gamma G\mu\left(t-t_b\right)\,,
\ee
$n\left(l,t\right)$ will have contributions from all pre-existing loops that have a physical length $l$ at a time $t$. So, the computation of the loop distribution function actually reduces to the computation of the time of formation of all the loops that contribute to it. In other words, we need to compute

\be
n\left(l,t\right)=\int_{t_i}^t dt_b\, \lp (l_b,t_b)\left(\frac{a(t_b)}{a(t)}\right)^3\,,
\ee
where we have taken into account the dilution of the loops caused by the background expansion, which yields

\be
n(l,t)=\sum_i\left\{\left(\alpha\left.\dtot{L}{t}\right|_{t=t_b^i}+\Gamma G\mu\right)^{-1}\frac{\cc\vv\left(t_b^i\right)}{\alpha L\left(t_b^i\right)^4}\left(\frac{a\left(t_b^i\right)}{a(t)}\right)^3\right\}\,,
\label{lnd-full}
\ee
where $t_b^i$ are the times of birth of loops that contribute to $n(l,t)$. Note that this expression does not rely on any assumption regarding cosmic string dynamics and it may be used throughout cosmological evolution (even during the radiation-matter transition and after the onset of dark-energy, when the network cannot maintain a linear scaling regime).

In order to illustrate the effect of having a distribution of loop lengths, instead of a single scale, at the moment of formation, let us assume that $a\propto t^\beta$ so that the network is in a linear scaling regime characterized by $L=\xi_{\beta}t$ and $\vv=v_\beta$ (where $\xi_\beta$ and $v_\beta$ are given by Eq.~(\ref{sca-par})). In this case, one has

\be
n(l,t)=\frac{A_\beta C_\beta(\alpha)}{t^{3\beta}\left(l+\Gamma G\mu t\right)^{4-3\beta}}\label{lndg}\,, 
\ee
where we have defined

\be
C_\beta(\alpha)=\frac{\left(\alpha\xi_\beta+\Gamma G\mu\right)^{3(\beta-1)}}{\alpha\xi_\beta} 
\ee
and

\be
A_\beta=\frac{\cc}{\sqrt{2}}\frac{v_\beta}{\xi_\beta ^3}\label{Ag}\,.
\ee

Let us now assume that the lengths of loops at the moment of production follow a given distribution (see also~\cite{Sanidas:2012ee,Blanco-Pillado:2013qja}). In principle, one may write the corresponding loop production function in terms of $\delta$-functions. Indeed, if one assumes that the energy lost as a result of loop production is distributed in $N$ different scales characterized by loop-size parameters $\alpha_1,\cdots,\alpha_N$, with associated weights $w_1,\cdots,w_N$, we should then have

\bq
\lp(l,t)=\sum_{i=1}^N w_i \frac{\cc}{\sqrt{2}l}\frac{\vv}{L^3}\delta(l-\alpha_i L)\equiv\sum_{i=1}^N w_i \lp_i(l,t)\\
\nonumber\xrightarrow[N\to\infty]{}\int w(\alpha')\lp_{\alpha'}(l,t)d\alpha'
\eq
Note that, in this case, we should have that

\be
\sum_{i=1}^Nw_i=1\xrightarrow[N\to\infty]{}\int w(\alpha')d\alpha'=1
\ee
in order to ensure energy balance (i.e. so as to have that all the energy lost by the network goes into the production of loops).

In this case, it is straightforward to show that

\be
n(l,t)=\sum_{i=1}^Nw_in_i(l,t)\xrightarrow[N\to\infty]{}\int d\alpha' w(\alpha')n_{\alpha'}(l,t)\,,
\ee
where $n_i(l,t)$ ($n_\alpha(l,t)$) is the loop production associated to the loop-size parameter $\alpha_i$ ($\alpha$). Using Eq.~(\ref{lndg}) one finds

\be
n(l,t)=\frac{\sum_{i=1}^Nw_iC_\beta(\alpha_i)A_\beta}{t^{3\beta}\left(l+\Gamma G\mu t\right)^{4-3\beta}}\xrightarrow[N\to\infty]{}\frac{\int d\alpha' w(\alpha')C_\beta(\alpha')A_\beta}{t^{3\beta}\left(l+\Gamma G\mu t\right)^{4-3\beta}}\,.\label{lnd-dist}
\ee
Note that for small enough $l$ (if we are not considering a distribution up to infinitely small $\alpha$), the result of the sum/integral in this expression is merely a number. Therefore, in this case, the effect of having a distribution of lengths at the moment of creation is to change the normalization of the loop production function (which can be computed if the exact distribution is known). For large $l$, however, only loops with $\alpha\gtrsim l/(\xi_\beta t)$ can contribute to $n(l,t)$ and the lower limit of the sum/integral is actually dependent on $l$ or $t$.

In any case, if the loop production function has a dominant scale $\alpha$ (i.e. if the distribution is peaked at a scale $\alpha$), one may in principle describe it using Eq.~(\ref{lndg}) provided that one corrects the normalization. In other words, one may redefine the parameter in Eq.~(\ref{Ag}) in such a way that

\be
A_\beta=\mathcal{F}\frac{\cc}{\sqrt{2}}\frac{v_\beta}{\xi_\beta^3}\,,
\ee
where $\mathcal{F}=(\int w(\alpha')C_\beta (\alpha')d\alpha')/C_\beta(\alpha)$ is a parameter that accounts for the spread of the distribution. As a matter of fact, it was shown in~\cite{Cui:2017ufi,Auclair:2019wcv} that the simulation-infered model of~\cite{Blanco-Pillado:2013qja} --- which exhibits loop production at a range of lengths --- is well described by this model for $\mathcal{F}=0.1$ and $l=0.1 t$.

For more complex distributions, with more than one very different (and prominent) loop production scale, this will lead to some loss of information in the low-frequency range of the spectrum: only one of the peaks of the spectrum, corresponding to the chosen dominant scale, will be predicted. Note however that Eq.~(\ref{lnd-dist})  also implies that

\be
\Ogw (f)=\sum_{i=1}^N w_i\Ogw(f,\alpha_i)\xrightarrow[N\to\infty]{}\int d\alpha' w_{\alpha'}\Ogw(f,\alpha')\,,
\ee
where $\Ogw(f,\alpha)$ is the spectrum obtained for a fixed loop-size parameter $\alpha$. Thus, for any known distribution of loops' lengths at the moment of formation, it is straightforward to construct the SGWB spectrum by using the analytical $\Ogw(f,\alpha)$. This simple analytical model may then be used to fully characterize the realistic SGWB generated by any (scaling) loop production function.

\section{Analytical approximation to the SGWB spectrum\label{sec-analytic}}

In order to make the problem of deriving an analytical approximation to the SGWB generated by cosmic string loops tractable, some simplifying assumptions are necessary. First of all, we will assume that the cosmic string network is in a linear scaling regime of the form of Eq.~(\ref{sca}) throughout its evolution, with $L=\xi_rt$ or $L=\xi_mt$ and $\vv=v_r$ or $\vv=v_m$ in the radiation or the matter era (with a sudden transition between the two regimes at the time of radiation-matter equality $t_{\rm eq}$). The parameters $\xi_r$($\xi_m$) and $v_r$($v_m$) are constant parameters given by Eqs.~(\ref{curv}) and~(\ref{sca-par}) with $\beta=1/2$ and $\beta=2/3$ respectively. As discussed in~\cite{Avelino:2012qy,Sousa:2013aaa}, this assumption is not realistic since the network cannot maintain scaling after the onset of the radiation-matter transition. As a result, this will lead to an underestimation of the amplitude of the contribution of the loops produced in the matter era to the SGWB. We will discuss the impact of this assumption in Sec.~\ref{sec-comp}. Moreover, we will consider two populations of loops: one created in the radiation era, with $t_i \le t_b\le t_{\rm eq}$, and another created in the matter era, with $t_{\rm eq} \le t_b \le t_0$. We will also assume that these two populations may be described by the same loop-size parameter $\alpha$: this is supported by the most recent (and largest) Nambu-Goto simulations~\cite{BlancoPillado:2011dq} of cosmic string networks, which found $\alpha_r \simeq 0.33$ and $\alpha_m \simeq 0.35$ in the radiation and matter eras respectively. Finally, we shall assume that the universe is effectively radiation(matter)-dominated, with $\beta=1/2\,(2/3)$, for $t\le t_{\rm eq}\,\left(t>t_{\rm eq}\right)$, but for consistency we shall assume that $H_0=100 h\,\mathrm{Km/s/Mpc}$, with $h=0.678$, and a flat universe in which the radiation and matter density parameters at the present time are respectively $\Omega_m=0.308$ and $\Omega_r=9.1476\times 10^{-5}$, as determined by Planck 2018 data~\cite{Aghanim:2018eyx}.

\subsection{Radiation era\label{subsec-rad}}

Let us start by considering the population of loops created in the radiation era, during which the Hubble parameter is effectively given by Eq.~(\ref{radera}). In this case, we have that

\be
n_r(l,t)=\frac{C_r(\alpha)}{t^{3/2}\left(l+\Gamma G\mu t\right)^{5/2}}\label{ldfr}\,, 
\ee
with

\be 
C_r(\alpha)=\mathcal{F}\frac{\cc v_r}{\sqrt{2}\alpha \xi_r^4}\left(\alpha\xi_r+\Gamma G\mu\right)^{3/2}\equiv \frac{A_r}{\alpha \xi_r}\left(\alpha\xi_r+\Gamma G\mu\right)^{3/2}\,.
\ee

Note that this expression is only valid for $t\le t_{\rm eq}$. However, if the loops are large, with $\alpha\ge\Gamma G\mu$, the loops created between $t_{\rm eq}/\left(1+\epsilon_r\right)<t_b< t_{\rm eq}$, with $\epsilon_r=\alpha\xi_r/(\Gamma G\mu$), are expected to survive beyond $t_{\rm eq}$ and to decay during the matter era. The number density of radiation-era loops in the matter era will merely result from the dilution and decay of the loops existing at the time of radiation-matter equality~\cite{Blanco-Pillado:2013qja}. Therefore, we have that

\be
n_{rm}(l,t)=\frac{C_r(\alpha)\left(2H_0\Omega_r^{1/2}\right)^{3/2}}{\left(l+\Gamma G\mu t\right)^{5/2}}\left(\frac{a_0}{a(t)}\right)^3\label{ldfrm}\,.
\ee

Let us start by considering the contribution of the radiation-era loops that decay during the radiation era. Any loop, created at a time $t_b$ will radiate GWs with frequencies larger than a minimum frequency $f_{min}$ defined as

\be
f\ge f_{min}(t_b)=\frac{2}{\alpha \xi_r t_b}\frac{a(t_b)}{a_0}\,. 
\ee
So, as time progresses, loops will contribute dominantly in progressively lower bins of frequency. As a result, the SGWB at a given frequency $f$ will only receive contributions from loops created after the instant of time $t_{min}$ for which $f_{min}(t_{min})=f$. In other words, one should only consider the contributions of loops created after

\be
\frac{a_0}{a_{min}}=\frac{\alpha\xi_r}{4H_0\Omega_r^{1/2}}f\label{minrad}\,, 
\ee
where $a_{min}=a(t_{min})$. The contribution of these loops to the SGWB spectrum is then

\be
\Ogw^r(f)=\frac{16\pi}{3}\left(\frac{G\mu}{H_0}\right)^2\frac{\Gamma}{fa_0}\int_{a_{min}}^{a_{\rm eq}} \left(\frac{a(t)}{a_0}\right)^4\frac{da}{H(a)}n_r(l,t)\label{Ogwr1}\,,
\ee
which yields

\be 
\Ogw^r(f)=\frac{128}{9}\pi A_r \Omega_r\frac{G\mu}{\epsilon_r}\left[\left(\frac{f(1+\epsilon_r)}{B_r\Omega_m/\Omega_r+f}\right)^{3/2}-1\right]\label{Ogwr2}\,,
\ee
where we have defined $B_r=4H_0\Omega_r^{1/2}/(\Gamma G\mu)$. Although at a first glance this expression looks different from the amplitude of the plateau we have derived in Eq.~(\ref{plateau}), the later is recovered in the limit $f\gg B_r\Omega_m/\Omega_r$.

As we have discussed earlier, the production of loops is only expected to become significant once the friction-dominated era of cosmic string dynamics ends, around $t_i\sim t_{\rm pl}/(G\mu)^2$. The SGWB spectrum will then necessarily have a cut-off at a certain frequency --- above which the spectrum should scale proportionally to $f^{-1}$~\cite{Guedes:2018afo} --- that is not included in Eq.~(\ref{Ogwr2}). This cut-off frequency may straightforwardly be computed by noting the following: since Eq.~(\ref{ldfr}) has the underlying assumption that the population of loops has already reached scaling (i.e., it assumes that there are loops in all stages of life, from ``birth'' to ``death", at the time $t$), strictly speaking it can only be considered valid after the first loop created by the network evaporates completely, for $t\ge(1+\epsilon_r)t_i$. Thus, the integration of Eq.~(\ref{Ogwr1}) should start at

\be
\frac{a_0}{a_{min}}=\frac{G\mu}{\left(2H_0\Omega_r^{1/2}(\epsilon_r+1)t_{\rm pl}\right)^{1/2}}\,.
\ee

As a result, the last term in Eq.~(\ref{Ogwr2}) should actually be

\be
\left(\frac{f(\epsilon_r+1)}{B_i+f}\right)^{3/2}\label{cut-off-full}\,,
\ee
with 

\be 
B_i=\frac{2}{\Gamma}\sqrt{\frac{2H_0\Omega^{1/2}}{t_{\rm pl}(\epsilon_r+1)}}\,,
\ee
instead of $1$. Note that, for $B_i\ll f$, this yields

\be
\left(\epsilon_r+1\right)^{3/2}\left[1-\frac{3}{2}\frac{B_i}{f}\right]\label{cut-off}
\ee
and, therefore, $f_{c}\sim 3B_i/2$ may be regarded as being approximately the frequency above which the spectrum has a cut-off.

The contribution of loops created and decaying in the radiation era in Eq.~(\ref{Ogwr2}), with the cut-off correction in Eq.~(\ref{cut-off-full}), is plotted in Fig.~\ref{conts} for $G\mu=10^{-10}$ and $\alpha=10^{-1}$ and $\alpha=10^{-5}$. Therein, one may see that, for $B_r\Omega_m/\Omega_r\ll f\ll B_i$, it does exhibit the expected characteristic plateau and that its amplitude decreases with decreasing loop size (approximately as $\sim \alpha^{1/2}$, in the large loop regime, as Eq.~(\ref{ldfr}) predicts). The $f^{-1}$ cut-off to the spectrum for frequencies above $f_c$ may also be observed there.
\\

As to the radiation-era loops that decay in the matter era, we need to compute

\be
\Ogw^{rm}(f)=\frac{16\pi}{3}\left(\frac{G\mu}{H_0}\right)^2\frac{\Gamma}{fa_0}\int_{a_{\rm eq}}^{a_{max}}\left(\frac{a(t)}{a_0}\right)^4\frac{da}{H(a)}n_{rm}(l,t)\,, 
\ee
and so one needs to find the last instant of time in which there is a contribution to the spectrum at the frequency $f$. At the time of radiation-matter equality, there are several loops with lengths in the range $0<l<\alpha\xi_r t_{\rm eq}$. However, for a given frequency $f$, only loops with a length $l\ge l_{min}(t_{\rm eq})=2a_{\rm eq}/(fa_0)$ will contribute to the SGWB. The loops with $l(t_{\rm eq})=l_{min}(t_{\rm eq})$ will be the first to contribute and it is simple to realize that the last ones would be those created at $t_{\rm eq}$. So this question reduces to finding the instant of time in which the loops created at $t_{\rm eq}$ contribute to the SGWB at a frequency $f$. Equivalently, we need to solve the equation

\be
\frac{f}{2}\left[\left(\alpha\xi_r+\Gamma G\mu\right)t_{\rm eq}-\Gamma G\mu t\right]=\frac{a(t)}{a_0}\label{amax}\,. 
\ee 
Note however that, since loops may survive for a significant period of time and since we are interested in computing $\Ogw^{rm}(f)$ at the present time, one should take $a_{max}=a_0$ if $a(t)$ satisfying Eq.~(\ref{amax}) exceeds $a_0$. As a matter of fact, the last loops created in the radiation era survive until $a_d=(\epsilon_r+1)^{3/2}a_{\rm eq}$ and large enough loops --- with $\epsilon_r\gtrsim (\Omega_m/\Omega_r)^{3/2}$ --- are expected to survive beyond the present time. So, except for frequencies $f\gg 2a_{\rm aeq}/(2\alpha\xi_rt_{\rm eq}a_0)$, one generally expects $a_{max}=a_0$ for large loops. Here, for simplicity, we will opt to always use $a_{max}=a_0$ for all $\epsilon_r\gtrsim 1$. Although this may lead to an over-estimation of this contribution, we shall see in Sec.~\ref{sec-comp} that this choice does not have a significant impact on the quality of the analytical approximation. Note also that since small loops, with $\epsilon_r\ll 1$, decay effectively immediately on the cosmological timescale~\cite{Sousa:2014gka}, radiation-era loops will not survive into the matter era in this limit. This contribution should not be included in this case.

So, by setting $a_{max}=a_0$ and assuming that the background is effectively matter-dominated, with

\be
H^2=H_0^2\Omega_m\left(\frac{a_0}{a(t)}\right)^3\label{hmat}\,, 
\ee
we find

\begin{widetext}
\bq
\Ogw^{rm}(f)=32\sqrt{3}\pi \left(\Omega_m\Omega_r\right)^{3/4}H_0\frac{A_r}{\Gamma}\frac{(\epsilon_r+1)^{3/2}}{f^{1/2}\epsilon_r}\left\{\frac{\left(\frac{\Omega_m}{\Omega_r}\right)^{1/4}}{\left(B_m\left(\frac{\Omega_m}{\Omega_r}\right)^{1/2}+f\right)^{1/2}}\left[2+\frac{f}{B_m\left(\frac{\Omega_m}{\Omega_r}\right)^{1/2}+f}\right]-\right.\label{Ogwrm}\\
\left.-\frac{1}{\left(B_m+f\right)^{1/2}}\left[2+\frac{f}{B_m+f}\right]\right\}\nonumber\,,
\eq
\end{widetext}
where we have defined $B_m=3H_0\Omega_m^{1/2}/(\Gamma G\mu)$. This contribution is also plotted in Fig.~\ref{conts}. As we may see therein, the radiation-era loops that decay in the matter era give rise to a sharp peak whose amplitude decreases as the length of loops decreases (also as $\alpha^{1/2}$ as expected).

\subsection{Matter era \label{subsec-mat}}

In the matter-dominated era, the loop number density is of the form

\be
n_m(l,t)=\frac{C_m(\alpha)}{t^2\left(l+\Gamma G\mu t\right)^2}\label{lndm}\,, 
\ee
with

\be
C_m(\alpha)=\mathcal{F}\frac{\cc}{\sqrt{2}\alpha}\frac{v_m}{\xi_m^4}\left(\alpha\xi_m+\Gamma G\mu\right)\equiv \frac{A_m}{\alpha\xi_m}\left(\alpha\xi_m+\Gamma G\mu\right)\,. 
\ee

We then need to compute

\be
\Ogw^m(f)=\frac{16\pi}{3}\left(\frac{G\mu}{H_0}\right)^2\frac{\Gamma}{fa_0}\int_{a_{min}}^{a_0}\left(\frac{a(t)}{a_0}\right)^4\frac{da}{H(a)}n_m(l,t)\,, 
\ee
where $a_{min}$ is the scale factor at the time of the production of the first matter-era loops that contribute to the SGWB. As in the case of the radiation-era loops, these loops should have $f_{min}=f$ and so

\be
\left(\frac{a_0}{a_{min}}\right)^{1/2}=\frac{\alpha\xi_m}{3H_0\Omega_m^{1/2}}f\,. 
\ee
Using Eqs.~(\ref{hmat}) and~(\ref{lndm}), we find

\begin{widetext}
\be 
\Ogw^m(f)=54\pi H_0\Omega_m^{3/2}\frac{A_m}{\Gamma}\frac{\epsilon_m+1}{\epsilon_m}\frac{B_m}{f}\left\{\frac{2B_m+f}{B_m(B_m+f)}-\frac{1}{f}\frac{2\epsilon_m+1}{\epsilon_m(\epsilon_m+1)}+\frac{2}{f}\log{\left(\frac{\epsilon_m+1}{\epsilon_m}\frac{B_m}{B_m+f}\right)}\right\}\,,
\label{Ogwm}
\ee
\end{widetext}
where we have introduced $\epsilon_m=\alpha\xi_m/(\Gamma G\mu)$. As we may see from this expression, the realistic picture is more complex than the simple scaling as $f^{-1}$ predicted in Eq.~(\ref{slope}). Note however that the different scalings that appear in this equation are a result of loop production ending at the present time, which introduces a cut-off in the low-frequency range. As a matter of fact, for $f\gg B_m$ (a range in which this sudden end of loop production is irrelevant), we have $\Ogw^m(f)\propto f^{-1}$. Note also that the production of matter-era loops only starts at $a_{\rm eq}$ and so, for large enough frequencies which have $a_{min}<a_{\rm eq}$, strictly speaking one should take $a_{min}=a_{\rm eq}$. However, since $\Ogw^m(f)\to\,0$ quickly as one increases the frequency, this only has a negligible effect on the spectrum.

\begin{figure}
\includegraphics[width=3.3in]{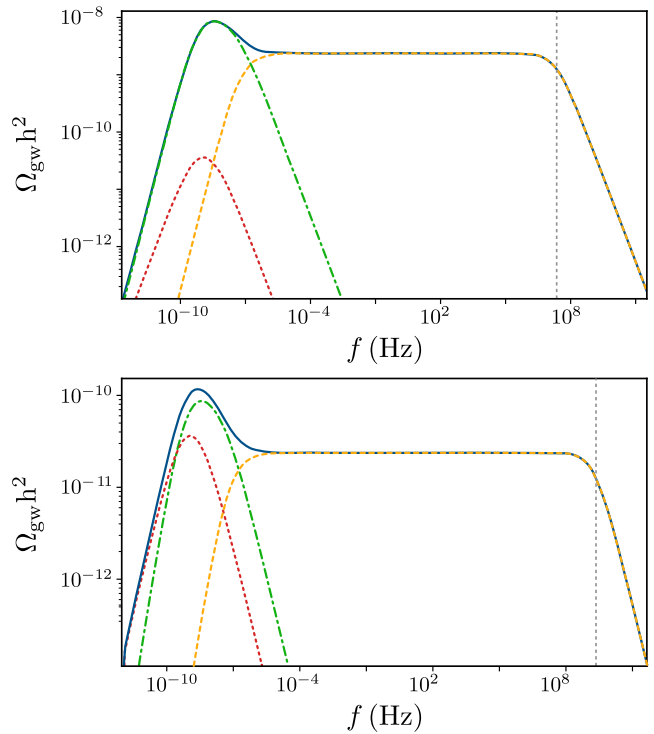}
\caption{Analytical approximation to the stochastic gravitational wave background generated by cosmic string networks with $G\mu=10^{-10}$ and $\alpha=10^{-1}$ (top panel) and $\alpha=10^{-5}$ (bottom panel). The solid (blue) lines represent the total SGWB, the dashed (yellow) lines represent the contribution of loops that are created and decay in the radiation era, the dash-dotted (green) lines corresponds to the radiation-era loops that decay during the matter era and the dotted (red) lines represents the contribution of loops created in the matter era. The vertical lines correspond to the cut-off frequency $f_c$.}
\label{conts}
\end{figure}

The contribution of the loops formed in the matter era to the stochastic gravitational wave spectrum is also represented in Fig.~\ref{conts}. Loops created in the matter era give rise to a sharply peaked spectrum (as the radiation-era loops that decay in the matter era do). However the peak of this contribution is located at a different frequency and the low-frequency cut-off scales more slowly with frequency. The actual shape and amplitude of the total SGWB spectrum will then be determined by the balance between the $\Ogw^{rm}(f)$ and $\Ogw^m(f)$ contributions. Therein we may also observe that, for large enough frequencies, the contribution from matter era loops to the SGWB does indeed scale proportionally to $f^{-1}$ as predicted in Eq.~(\ref{slope}). More interestingly, though, this contribution is, in the large loop regime, roughly independent of $\alpha$ (cf. Eq.~(\ref{lndm})) and therefore its maximum amplitude remains effectively unchanged as we decrease $\alpha$. As a result, if the loop-size parameter is decreased, the relative importance of the radiation-era loops that decay in the matter era also decreases when compared to that of matter-era loops. So, the effect of the choice $a_{max}=a_0$ for all $\alpha$ is mitigated even when $a_{max}<a_0$ for $\epsilon_r<(\Omega_m/\Omega_r)^{3/2}$.

\subsection{Small loop regime\label{subsec-small}}

Although the expressions derived in the previous sections also apply to small loops, with $\alpha<\Gamma G\mu$ (provided one switches off the $\Ogw^{rm}(f)$ contribution), the fact that loops decay effectively immediately after creation allows us to further simplify this approximation. In this case, one has that~\cite{Sousa:2014gka}

\be
\Ogw(f)=\frac{16\pi}{3}\frac{G}{H_0^2}\frac{1}{f}\int_{t_{\rm min}}^{t_0}\left.\dtot{\rho}{t}\right|_{\rm loops}\frac{dt}{\alpha L(t)}\left(\frac{a(t)}{a_0}\right)^5\,,
\ee
and thus
\be
\Ogw(f)=\frac{64\pi}{3} G\mu \Omega_r A_r+54\pi\frac{H_0\Omega_m^{3/2}}{\epsilon_m\Gamma}\frac{A_m}{f}\left[1-\frac{B_m}{\epsilon_m}\frac{1}{f}\right]\,,
\label{approx-small}
\ee
which is equivalent to the approximation found in~\cite{Sousa:2014gka} for the small-loop regime.

\section{Comparison with spectra obtained numerically\label{sec-comp}}

To assess the quality of the analytical approximation derived in the previous section, we will compare our results with those obtained by integrating Eq.~(\ref{spec}) in a realistic cosmological background (which is assumed to contain radiation, matter and a cosmological constant). When computing these spectra numerically, we shall make minimal simplifying assumptions regarding cosmic string dynamics --- i.e. we will solve Eqs.~(\ref{eqvosv}) and~(\ref{eqvosL}) (together with the Friedmann equation) throughout cosmological history --- and we will determine $n(l,t)$ numerically using Eq.~(\ref{lnd-full}) to accurately gauge it through the radiation-matter transition as in Ref.~\cite{Sousa:2013aaa}.

\begin{figure}
\includegraphics[width=3.3in]{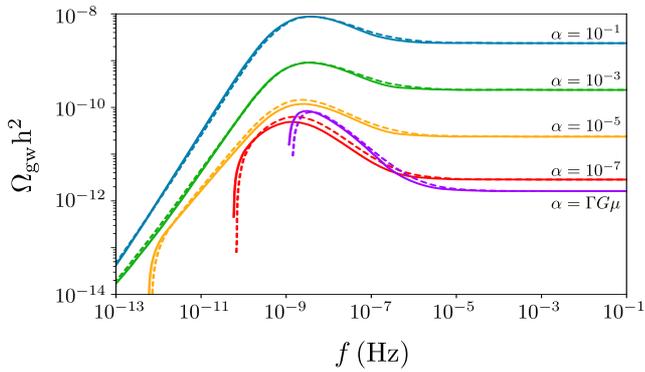}
\caption{Analytical approximation to the stochastic gravitational wave  background generated by cosmic string networks with $G\mu=10^{-10}$ and different values of $\alpha$. The solid lines represent the approximation to the SGWB in Eq.~(\ref{approx-large}), while the dashed lines correspond to the SGWB obtained numerically.}
\label{comp-large}
\end{figure}

\begin{figure}
\includegraphics[width=3.3in]{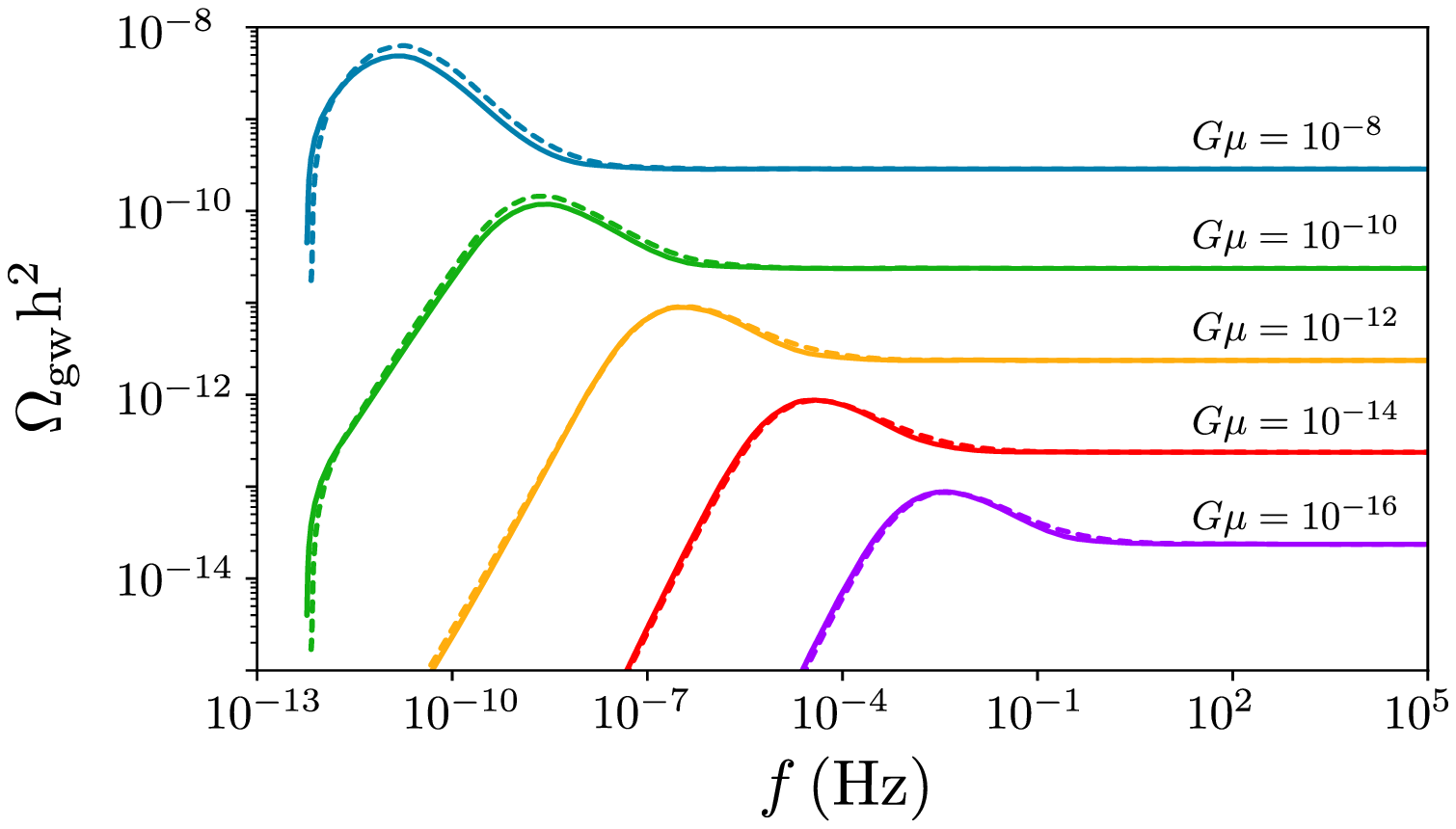}
\caption{Analytical approximation to the stochastic gravitational wave  background generated by cosmic string networks with $\alpha=10^{-5}$ and different values of $G\mu$. The solid lines represent the approximation to the SGWB in Eq.~(\ref{approx-large}), while the dashed lines correspond to the SGWB obtained numerically.}
\label{comp-gmus}
\end{figure}

Let us start by considering the case in which the loops produced by the network are large, with $\epsilon_r,\epsilon_m\gg1$. In this case, the spectrum is approximated by

\be
\Ogw(f)=\Ogw^r(f)+\Ogw^{rm}(f)+\Ogw^m(f)\,,
\label{approx-large}
\ee
where $\Ogw^r(f)$, $\Ogw^{rm}(f)$ and $\Ogw^m(f)$ are given by Eqs.~(\ref{Ogwr2}),~(\ref{Ogwrm}) and~(\ref{Ogwm}), respectively. In Fig.~\ref{comp-large}, the approximation in Eq.~(\ref{approx-large}) is plotted alongside spectra  obtained numerically for $G\mu=10^{-10}$ and different values of $\alpha$. As this figure illustrates, this approximation provides a very good description of the realistic SGWB generated by cosmic strings throughout the observationally relevant frequency range. This approximation is exact in the high-frequency range corresponding to the radiation-era plateau. However, some discrepancies can be observed in the low-frequency range, which is where the (necessary) simplifying assumptions can have some impact on our results. First of all, the peaks of the approximated spectra are, in general, slightly less broad than the peaks of the spectra obtained numerically. This is a consequence of assuming that the radiation-matter transition happens abruptly at $t=t_{\rm eq}$, instead of the expected smooth transition. Moreover, in the very low-frequency range, one may see that there is a small overestimation of the amplitude of the spectrum, since we did not account for the dilution of cosmic strings and string loops caused by the recent onset of dark energy in our numerical approximation. This effect may be seen in Fig.~\ref{comp-large} in the spectra with $\alpha\le 10^{-5}$, but it is also present for larger $\alpha$. More important, though, is the effect of assuming that the cosmic string network remains in a linear scaling regime throughout its evolution. As discussed in~\cite{Sousa:2013aaa}, this leads to a significant underestimation of the number of loops produced in the matter era and, consequently, to an underestimation of the amplitude of the spectrum. It is precisely this effect that causes the discrepancies in the amplitude of the peaks of the spectra plotted in Fig.~\ref{comp-large}. For large $\alpha$, this effect is barely visible since the dominant contribution in the frequency range corresponding to the peak is that of the loops created in the radiation era that survive into the matter era, $\Ogw^{rm}(f)$ (although this effect is, in fact, present at lower frequencies~\cite{Auclair:2019wcv}). However, as Fig.~\ref{conts} illustrates, as we lower $\alpha$, $\Ogw^m(f)$ becomes more relevant to the peak of the spectrum and eventually dominates the contribution in this frequency range. As a result, and since the assumption of linear scaling leads to an underestimation of this contribution, the quality of our approximation is poorer for lower $\alpha$. The same can been seen in Fig.~\ref{comp-gmus}, where we plot the SGWB generated by cosmic strings for fixed $\alpha=10^{-5}$ for different values of $G\mu$ alongside the analytical approximation in Eq.~(\ref{approx-large}): the larger $\alpha$ is when compared to $G\mu$, the better our approximation is.

\begin{figure}
\includegraphics[width=3.3in]{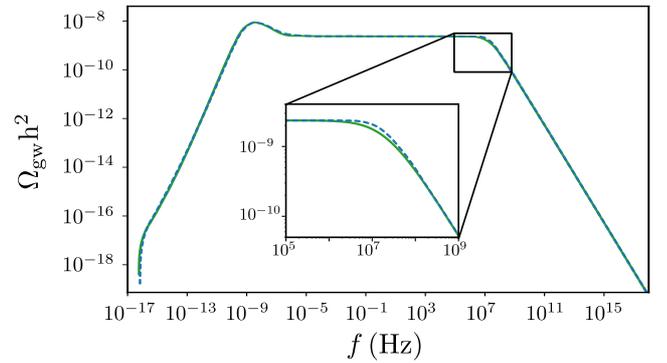}
\caption{Analytical approximation to the full stochastic gravitational wave  background generated by cosmic string networks with $\alpha=10^{-1}$ and $G\mu=10^{-10}$, including the cut-off correction in Eq.~(\ref{cut-off-full}). The solid line represents the approximation to the SGWB in Eq.~(\ref{approx-large}), while the dashed line corresponds to the SGWB obtained numerically.}
\label{comp-cutoff}
\end{figure}

Note, however, that in any case this approximation is generally very good: the maximum relative difference in the amplitude of the peak of the spectrum never exceeds $30\%$ in the large loop limit and is, in general, significantly smaller than this value. It may be used to describe with minor error only the full SGWB spectrum generated by cosmic string networks, if one includes the correction in Eq.~(\ref{cut-off-full}). We plot an example of the approximation to the full SGWB spectrum, including the correction in Eq.~(\ref{cut-off-full}), with $G\mu=10^{-10}$ and $\alpha=10^{-1}$ in Fig.~\ref{comp-cutoff}. Therein, we may see that this correction indeed provides a good description of the high-frequency cut-off of the spectrum, albeit with a slight underestimation of the amplitude of the spectrum in the transitional region between the plateau and the cut-off (that is caused by the fact that, when deriving the approximation, we only considered the contribution of loops once scaling is reached). We have verified numerically that $\Ogw^r(f)$ with the cut-off correction in Eq.~(\ref{cut-off-full}) provides a good fit to the spectrum down to its cut-off for $\alpha\ge 10\,\Gamma G\mu$. For smaller $\alpha$, if one is interested in describing the spectrum up to frequencies $f>f_c\sim 3.5\times 10^{10}/(1+\epsilon_r)\,\,{\rm Hz}$ (which is currently outside of the sensitivity windows of the major current and upcoming GW experiments~\cite{Guedes:2018afo}), one may use Eq.~(\ref{Ogwr2}) without any correction for $f<f_c$ and manually switch on the correction in Eq.~(\ref{cut-off}) for $f>f_c$ (which always works in the $f\gg f_c$ limit).

\begin{figure}
\includegraphics[width=3.3in]{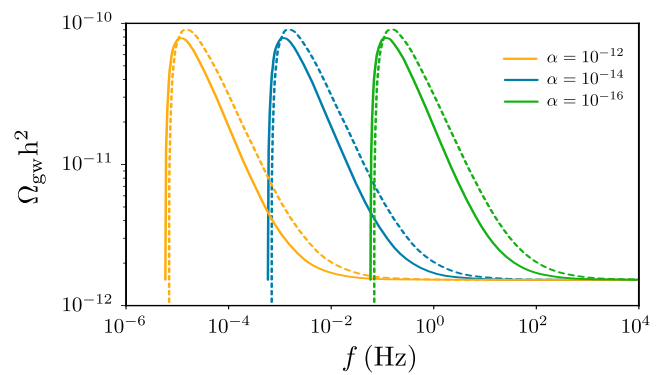}
\caption{Analytical approximation to the stochastic gravitational wave  background generated by cosmic string networks with $G\mu=10^{-10}$ for different values of $\alpha$ in the small loop regime. The solid lines represent the approximation to the SGWB in Eq.~(\ref{approx-small}), while the dashed lines correspond to the SGWB obtained numerically.}
\label{comp-small}
\end{figure}

In the small-loop regime, with $\alpha\ll\Gamma G\mu$, as we have seen, our approximation may be reduced to Eq.~(\ref{approx-small}). We have verified numerically that this approximation provides a good fit to the numerical spectrum --- with the discrepancies caused by our assumptions discussed for the large loop regime --- for $\alpha<0.1\Gamma G\mu$, as shown in Fig.~(\ref{comp-small}). However, the small-loop regime is precisely the limit in which the underestimation of the matter-era contribution is more dramatic and this is patent in this figure. The maximum error in the height of the peak of the spectrum is approximately $45\%$ in this case. Note that in the range $\Gamma G\mu>\alpha\ge 0.1\Gamma G\mu$ one still has to resort to Eq.~(\ref{approx-large}), but without the contribution $\Ogw^{rm}(f)$.

\begin{figure}
\includegraphics[width=3.3in]{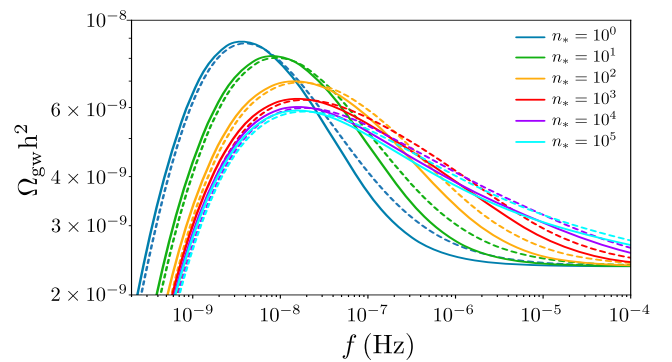}
\caption{Analytical approximation to the stochastic gravitational wave  background generated by cosmic string networks with $\alpha=10^{-1}$ and $G\mu=10^{-10}$, $q=4/3$ and different values of $n_*$. The solid lines represent the approximation to the SGWB, while the dashed lines correspond to the SGWB obtained numerically.}
\label{comp-high}
\end{figure}

Finally, up until now we have only considered the contribution of the fundamental mode of emission to the SGWB generated by cosmic string networks. The inclusion of higher harmonic modes, however, is expected to have a significant impact on the spectrum and may significantly alter its shape. Nevertheless, one may construct an analytical approximation up to an arbitrary number of modes of emission $n_*$, for any $q$, by using Eqs.~(\ref{higher1}) and~(\ref{higher2}):

\be
\Ogw(f,q,n_*)=\sum_{j=1}^{n_*}\frac{j^{-q}}{\mathcal{E}}\Ogw^1(f/j)\,.
\ee
As Fig.~\ref{comp-high} illustrates, for $G\mu=10^{-10}$, $\alpha=10^{-1}$, $q=4/3$ and various values of $n_*$, the quality of the analytical approximation is, as expected, not significantly affected by the inclusion of higher emission modes.

\section{Conclusions\label{sec-conclusions}}

We have derived a full analytical approximation to the SGWB  generated by cosmic string networks as a function of $\alpha$, $G  \mu$, $q$, $n_*$, significantly extending previous work, which was  restricted to the limit where the loops produced by the networks  are small. By comparing it with the numerical results  obtained taking into account the full cosmological network dynamics  as given by the cosmic string VOS model, we have demonstrated that  this analytical result provides an excellent approximation to the  SGWB power spectrum generated by standard cosmic strings. Our  analytical approximation has been shown to provide a high quality  fit to the SGWB power spectrum for a wide range of frequencies and  for the relevant ($\alpha$, $G \mu$, $q$, $n_*$) parameter space,  with the maximum relative error in the height of the peak of the  spectrum never exceeding $50 \%$ both in the small and large loop  regimes --- and being, in general, significantly smaller than that. Note that, although this approximation was derived under the assumption that loop production occurs at a single lengthscale, our results can, in principle, be used to construct an analytical approximation to more complex scenarios in which the length of loops at moment of creation follows a distribution (as demonstrated by the results in Sec.~\ref{sec-ldf}).  Moreover, although the results presented here correspond to standard strings, which are well described by the VOS model with $\cc=0.23$, this approximation may, in principle, also be used to characterize the SGWB generated by non-standard networks with different values of $\cc$ provided there is significant loop production (for instance, strings which have a reduced intercommutation probability~\cite{Sousa:2016ggw}).

This analytical fit is expected to become an extremely useful tool  for fast estimations of the SGWB power spectrum generated by cosmic  string networks over a wide parameter space, thus allowing for a  quick derivation of the observational constraints on cosmic string  scenarios from current and forthcoming GW experiments.

\acknowledgments
L.S. thanks Ivan Rybak and the members of the Cosmology Working Group of the LISA Consortium for interesting discussions. L. S. is supported by FCT - Funda\c{c}\~{a}o para a Ci\^{e}ncia e a Tecnologia through contract No. DL 57/2016/CP1364/CT0001. P. P. A. thanks the support from FCT---Funda{\c c}\~ao para a Ci\^encia e a Tecnologia through the Sabbatical Grant No. SFRH/BSAB/150322/2019. G. G. is supported by FCT under the grant SFRH/BD/144244/2019 and also acknowledges financial support from LIP (UID/FIS/50007/2019, FCT, COMPETE2020-Portugal2020, FEDER, POCI-01-0145-FEDER-007334). This work was also supported by FCT through national funds (PTDC/FIS-PAR/31938/2017) and by FEDER - Fundo Europeu de Desenvolvimento Regional through COMPETE2020 - Programa Operacional Competitividade e Internacionaliza\c{c}\~{a}o (POCI-01-0145-FEDER-031938). This work was also supported by  FCT through the research grants UID/FIS/04434/2019, UIDB/04434/2020 and UIDP/04434/2020.

\bibliography{approx}

\begin{thebibliography}{43}
\expandafter\ifx\csname natexlab\endcsname\relax\def\natexlab#1{#1}\fi
\expandafter\ifx\csname bibnamefont\endcsname\relax
  \def\bibnamefont#1{#1}\fi
\expandafter\ifx\csname bibfnamefont\endcsname\relax
  \def\bibfnamefont#1{#1}\fi
\expandafter\ifx\csname citenamefont\endcsname\relax
  \def\citenamefont#1{#1}\fi
\expandafter\ifx\csname url\endcsname\relax
  \def\url#1{\texttt{#1}}\fi
\expandafter\ifx\csname urlprefix\endcsname\relax\def\urlprefix{URL }\fi
\providecommand{\bibinfo}[2]{#2}
\providecommand{\eprint}[2][]{\url{#2}}

\bibitem[{\citenamefont{Kibble}(1976)}]{Kibble:1976sj}
\bibinfo{author}{\bibfnamefont{T.~W.~B.} \bibnamefont{Kibble}},
  \bibinfo{journal}{J. Phys.} \textbf{\bibinfo{volume}{A9}},
  \bibinfo{pages}{1387} (\bibinfo{year}{1976}).

\bibitem[{\citenamefont{Hindmarsh and Kibble}(1995)}]{Hindmarsh:1994re}
\bibinfo{author}{\bibfnamefont{M.~B.} \bibnamefont{Hindmarsh}}
  \bibnamefont{and} \bibinfo{author}{\bibfnamefont{T.~W.~B.}
  \bibnamefont{Kibble}}, \bibinfo{journal}{Rept. Prog. Phys.}
  \textbf{\bibinfo{volume}{58}}, \bibinfo{pages}{477} (\bibinfo{year}{1995}),
  \eprint{hep-ph/9411342}.

\bibitem[{\citenamefont{Vilenkin and Shellard}(2000)}]{Vilenkin:2000jqa}
\bibinfo{author}{\bibfnamefont{A.}~\bibnamefont{Vilenkin}} \bibnamefont{and}
  \bibinfo{author}{\bibfnamefont{E.~P.~S.} \bibnamefont{Shellard}},
  \emph{\bibinfo{title}{{Cosmic Strings and Other Topological Defects}}}
  (\bibinfo{publisher}{Cambridge University Press}, \bibinfo{year}{2000}), ISBN
  \bibinfo{isbn}{9780521654760}.

\bibitem[{\citenamefont{Vilenkin}(1981)}]{Vilenkin:1981bx}
\bibinfo{author}{\bibfnamefont{A.}~\bibnamefont{Vilenkin}},
  \bibinfo{journal}{Phys. Lett.} \textbf{\bibinfo{volume}{107B}},
  \bibinfo{pages}{47} (\bibinfo{year}{1981}).

\bibitem[{\citenamefont{Hogan and Rees}(1984)}]{Hogan:1984is}
\bibinfo{author}{\bibfnamefont{C.~J.} \bibnamefont{Hogan}} \bibnamefont{and}
  \bibinfo{author}{\bibfnamefont{M.~J.} \bibnamefont{Rees}},
  \bibinfo{journal}{Nature} \textbf{\bibinfo{volume}{311}},
  \bibinfo{pages}{109} (\bibinfo{year}{1984}), \bibinfo{note}{[,128(1984)]}.

\bibitem[{\citenamefont{Accetta and Krauss}(1989)}]{Accetta:1988bg}
\bibinfo{author}{\bibfnamefont{F.~S.} \bibnamefont{Accetta}} \bibnamefont{and}
  \bibinfo{author}{\bibfnamefont{L.~M.} \bibnamefont{Krauss}},
  \bibinfo{journal}{Nucl. Phys.} \textbf{\bibinfo{volume}{B319}},
  \bibinfo{pages}{747} (\bibinfo{year}{1989}).

\bibitem[{\citenamefont{Abbott et~al.}(2016)}]{Abbott:2016nmj}
\bibinfo{author}{\bibfnamefont{B.~P.} \bibnamefont{Abbott}}
  \bibnamefont{et~al.} (\bibinfo{collaboration}{LIGO Scientific, Virgo}),
  \bibinfo{journal}{Phys. Rev. Lett.} \textbf{\bibinfo{volume}{116}},
  \bibinfo{pages}{241103} (\bibinfo{year}{2016}), \eprint{1606.04855}.

\bibitem[{\citenamefont{Abbott et~al.}(2017)}]{TheLIGOScientific:2017qsa}
\bibinfo{author}{\bibfnamefont{B.~P.} \bibnamefont{Abbott}}
  \bibnamefont{et~al.} (\bibinfo{collaboration}{LIGO Scientific, Virgo}),
  \bibinfo{journal}{Phys. Rev. Lett.} \textbf{\bibinfo{volume}{119}},
  \bibinfo{pages}{161101} (\bibinfo{year}{2017}), \eprint{1710.05832}.

\bibitem[{\citenamefont{Auclair et~al.}(2019{\natexlab{a}})}]{Auclair:2019wcv}
\bibinfo{author}{\bibfnamefont{P.}~\bibnamefont{Auclair}} \bibnamefont{et~al.}
  (\bibinfo{year}{2019}{\natexlab{a}}), \eprint{1909.00819}.

\bibitem[{\citenamefont{Lentati et~al.}(2015)}]{Lentati:2015qwp}
\bibinfo{author}{\bibfnamefont{L.}~\bibnamefont{Lentati}} \bibnamefont{et~al.},
  \bibinfo{journal}{Mon. Not. Roy. Astron. Soc.}
  \textbf{\bibinfo{volume}{453}}, \bibinfo{pages}{2576} (\bibinfo{year}{2015}),
  \eprint{1504.03692}.

\bibitem[{\citenamefont{Sanidas et~al.}(2012)\citenamefont{Sanidas, Battye, and
  Stappers}}]{Sanidas:2012ee}
\bibinfo{author}{\bibfnamefont{S.~A.} \bibnamefont{Sanidas}},
  \bibinfo{author}{\bibfnamefont{R.~A.} \bibnamefont{Battye}},
  \bibnamefont{and} \bibinfo{author}{\bibfnamefont{B.~W.}
  \bibnamefont{Stappers}}, \bibinfo{journal}{Phys. Rev.}
  \textbf{\bibinfo{volume}{D85}}, \bibinfo{pages}{122003}
  (\bibinfo{year}{2012}), \eprint{1201.2419}.

\bibitem[{\citenamefont{Binetruy et~al.}(2012)\citenamefont{Binetruy, Bohe,
  Caprini, and Dufaux}}]{Binetruy:2012ze}
\bibinfo{author}{\bibfnamefont{P.}~\bibnamefont{Binetruy}},
  \bibinfo{author}{\bibfnamefont{A.}~\bibnamefont{Bohe}},
  \bibinfo{author}{\bibfnamefont{C.}~\bibnamefont{Caprini}}, \bibnamefont{and}
  \bibinfo{author}{\bibfnamefont{J.-F.} \bibnamefont{Dufaux}},
  \bibinfo{journal}{JCAP} \textbf{\bibinfo{volume}{1206}}, \bibinfo{pages}{027}
  (\bibinfo{year}{2012}), \eprint{1201.0983}.

\bibitem[{\citenamefont{Kuroyanagi et~al.}(2012)\citenamefont{Kuroyanagi,
  Miyamoto, Sekiguchi, Takahashi, and Silk}}]{Kuroyanagi:2012wm}
\bibinfo{author}{\bibfnamefont{S.}~\bibnamefont{Kuroyanagi}},
  \bibinfo{author}{\bibfnamefont{K.}~\bibnamefont{Miyamoto}},
  \bibinfo{author}{\bibfnamefont{T.}~\bibnamefont{Sekiguchi}},
  \bibinfo{author}{\bibfnamefont{K.}~\bibnamefont{Takahashi}},
  \bibnamefont{and} \bibinfo{author}{\bibfnamefont{J.}~\bibnamefont{Silk}},
  \bibinfo{journal}{Phys. Rev.} \textbf{\bibinfo{volume}{D86}},
  \bibinfo{pages}{023503} (\bibinfo{year}{2012}), \eprint{1202.3032}.

\bibitem[{\citenamefont{Sousa and Avelino}(2013)}]{Sousa:2013aaa}
\bibinfo{author}{\bibfnamefont{L.}~\bibnamefont{Sousa}} \bibnamefont{and}
  \bibinfo{author}{\bibfnamefont{P.~P.} \bibnamefont{Avelino}},
  \bibinfo{journal}{Phys. Rev.} \textbf{\bibinfo{volume}{D88}},
  \bibinfo{pages}{023516} (\bibinfo{year}{2013}), \eprint{1304.2445}.

\bibitem[{\citenamefont{Sousa and Avelino}(2016)}]{Sousa:2016ggw}
\bibinfo{author}{\bibfnamefont{L.}~\bibnamefont{Sousa}} \bibnamefont{and}
  \bibinfo{author}{\bibfnamefont{P.~P.} \bibnamefont{Avelino}},
  \bibinfo{journal}{Phys. Rev.} \textbf{\bibinfo{volume}{D94}},
  \bibinfo{pages}{063529} (\bibinfo{year}{2016}), \eprint{1606.05585}.

\bibitem[{\citenamefont{Blanco-Pillado and
  Olum}(2017)}]{Blanco-Pillado:2017oxo}
\bibinfo{author}{\bibfnamefont{J.~J.} \bibnamefont{Blanco-Pillado}}
  \bibnamefont{and} \bibinfo{author}{\bibfnamefont{K.~D.} \bibnamefont{Olum}},
  \bibinfo{journal}{Phys. Rev.} \textbf{\bibinfo{volume}{D96}},
  \bibinfo{pages}{104046} (\bibinfo{year}{2017}), \eprint{1709.02693}.

\bibitem[{\citenamefont{Ringeval and Suyama}(2017)}]{Ringeval:2017eww}
\bibinfo{author}{\bibfnamefont{C.}~\bibnamefont{Ringeval}} \bibnamefont{and}
  \bibinfo{author}{\bibfnamefont{T.}~\bibnamefont{Suyama}},
  \bibinfo{journal}{JCAP} \textbf{\bibinfo{volume}{1712}}, \bibinfo{pages}{027}
  (\bibinfo{year}{2017}), \eprint{1709.03845}.

\bibitem[{\citenamefont{Cui et~al.}(2019)\citenamefont{Cui, Lewicki, Morrissey,
  and Wells}}]{Cui:2018rwi}
\bibinfo{author}{\bibfnamefont{Y.}~\bibnamefont{Cui}},
  \bibinfo{author}{\bibfnamefont{M.}~\bibnamefont{Lewicki}},
  \bibinfo{author}{\bibfnamefont{D.~E.} \bibnamefont{Morrissey}},
  \bibnamefont{and} \bibinfo{author}{\bibfnamefont{J.~D.} \bibnamefont{Wells}},
  \bibinfo{journal}{JHEP} \textbf{\bibinfo{volume}{01}}, \bibinfo{pages}{081}
  (\bibinfo{year}{2019}), \eprint{1808.08968}.

\bibitem[{\citenamefont{Jenkins and Sakellariadou}(2018)}]{Jenkins:2018lvb}
\bibinfo{author}{\bibfnamefont{A.~C.} \bibnamefont{Jenkins}} \bibnamefont{and}
  \bibinfo{author}{\bibfnamefont{M.}~\bibnamefont{Sakellariadou}},
  \bibinfo{journal}{Phys. Rev.} \textbf{\bibinfo{volume}{D98}},
  \bibinfo{pages}{063509} (\bibinfo{year}{2018}), \eprint{1802.06046}.

\bibitem[{\citenamefont{Auclair
  et~al.}(2019{\natexlab{b}})\citenamefont{Auclair, Steer, and
  Vachaspati}}]{Auclair:2019jip}
\bibinfo{author}{\bibfnamefont{P.}~\bibnamefont{Auclair}},
  \bibinfo{author}{\bibfnamefont{D.~A.} \bibnamefont{Steer}}, \bibnamefont{and}
  \bibinfo{author}{\bibfnamefont{T.}~\bibnamefont{Vachaspati}}
  (\bibinfo{year}{2019}{\natexlab{b}}), \eprint{1911.12066}.

\bibitem[{\citenamefont{Gouttenoire et~al.}(2019)\citenamefont{Gouttenoire,
  Servant, and Simakachorn}}]{Gouttenoire:2019kij}
\bibinfo{author}{\bibfnamefont{Y.}~\bibnamefont{Gouttenoire}},
  \bibinfo{author}{\bibfnamefont{G.}~\bibnamefont{Servant}}, \bibnamefont{and}
  \bibinfo{author}{\bibfnamefont{P.}~\bibnamefont{Simakachorn}}
  (\bibinfo{year}{2019}), \eprint{1912.02569}.

\bibitem[{\citenamefont{Martins and Shellard}(1996)}]{Martins:1996jp}
\bibinfo{author}{\bibfnamefont{C.~J. A.~P.} \bibnamefont{Martins}}
  \bibnamefont{and} \bibinfo{author}{\bibfnamefont{E.~P.~S.}
  \bibnamefont{Shellard}}, \bibinfo{journal}{Phys. Rev.}
  \textbf{\bibinfo{volume}{D54}}, \bibinfo{pages}{2535} (\bibinfo{year}{1996}),
  \eprint{hep-ph/9602271}.

\bibitem[{\citenamefont{Martins and Shellard}(2002)}]{Martins:2000cs}
\bibinfo{author}{\bibfnamefont{C.~J. A.~P.} \bibnamefont{Martins}}
  \bibnamefont{and} \bibinfo{author}{\bibfnamefont{E.~P.~S.}
  \bibnamefont{Shellard}}, \bibinfo{journal}{Phys. Rev.}
  \textbf{\bibinfo{volume}{D65}}, \bibinfo{pages}{043514}
  (\bibinfo{year}{2002}), \eprint{hep-ph/0003298}.

\bibitem[{\citenamefont{Kibble}(1985)}]{Kibble:1984hp}
\bibinfo{author}{\bibfnamefont{T.~W.~B.} \bibnamefont{Kibble}},
  \bibinfo{journal}{Nucl. Phys.} \textbf{\bibinfo{volume}{B252}},
  \bibinfo{pages}{227} (\bibinfo{year}{1985}), \bibinfo{note}{[Erratum: Nucl.
  Phys.B261,750(1985)]}.

\bibitem[{\citenamefont{Burden}(1985)}]{Burden:1985md}
\bibinfo{author}{\bibfnamefont{C.~J.} \bibnamefont{Burden}},
  \bibinfo{journal}{Phys. Lett.} \textbf{\bibinfo{volume}{164B}},
  \bibinfo{pages}{277} (\bibinfo{year}{1985}).

\bibitem[{\citenamefont{Allen and Shellard}(1992)}]{Allen:1991bk}
\bibinfo{author}{\bibfnamefont{B.}~\bibnamefont{Allen}} \bibnamefont{and}
  \bibinfo{author}{\bibfnamefont{E.~P.~S.} \bibnamefont{Shellard}},
  \bibinfo{journal}{Phys. Rev.} \textbf{\bibinfo{volume}{D45}},
  \bibinfo{pages}{1898} (\bibinfo{year}{1992}).

\bibitem[{\citenamefont{Vachaspati and Vilenkin}(1985)}]{Vachaspati:1984gt}
\bibinfo{author}{\bibfnamefont{T.}~\bibnamefont{Vachaspati}} \bibnamefont{and}
  \bibinfo{author}{\bibfnamefont{A.}~\bibnamefont{Vilenkin}},
  \bibinfo{journal}{Phys. Rev.} \textbf{\bibinfo{volume}{D31}},
  \bibinfo{pages}{3052} (\bibinfo{year}{1985}).

\bibitem[{\citenamefont{Binetruy et~al.}(2009)\citenamefont{Binetruy, Bohe,
  Hertog, and Steer}}]{Binetruy:2009vt}
\bibinfo{author}{\bibfnamefont{P.}~\bibnamefont{Binetruy}},
  \bibinfo{author}{\bibfnamefont{A.}~\bibnamefont{Bohe}},
  \bibinfo{author}{\bibfnamefont{T.}~\bibnamefont{Hertog}}, \bibnamefont{and}
  \bibinfo{author}{\bibfnamefont{D.~A.} \bibnamefont{Steer}},
  \bibinfo{journal}{Phys. Rev.} \textbf{\bibinfo{volume}{D80}},
  \bibinfo{pages}{123510} (\bibinfo{year}{2009}), \eprint{0907.4522}.

\bibitem[{\citenamefont{Scherrer et~al.}(1990)\citenamefont{Scherrer,
  Quashnock, Spergel, and Press}}]{Scherrer:1990pj}
\bibinfo{author}{\bibfnamefont{R.~J.} \bibnamefont{Scherrer}},
  \bibinfo{author}{\bibfnamefont{J.~M.} \bibnamefont{Quashnock}},
  \bibinfo{author}{\bibfnamefont{D.~N.} \bibnamefont{Spergel}},
  \bibnamefont{and} \bibinfo{author}{\bibfnamefont{W.~H.} \bibnamefont{Press}},
  \bibinfo{journal}{Phys. Rev.} \textbf{\bibinfo{volume}{D42}},
  \bibinfo{pages}{1908} (\bibinfo{year}{1990}).

\bibitem[{\citenamefont{Quashnock and Spergel}(1990)}]{Quashnock:1990wv}
\bibinfo{author}{\bibfnamefont{J.~M.} \bibnamefont{Quashnock}}
  \bibnamefont{and} \bibinfo{author}{\bibfnamefont{D.~N.}
  \bibnamefont{Spergel}}, \bibinfo{journal}{Phys. Rev.}
  \textbf{\bibinfo{volume}{D42}}, \bibinfo{pages}{2505} (\bibinfo{year}{1990}).

\bibitem[{\citenamefont{Brandenberger et~al.}(1986)\citenamefont{Brandenberger,
  Albrecht, and Turok}}]{Brandenberger:1986xn}
\bibinfo{author}{\bibfnamefont{R.~H.} \bibnamefont{Brandenberger}},
  \bibinfo{author}{\bibfnamefont{A.}~\bibnamefont{Albrecht}}, \bibnamefont{and}
  \bibinfo{author}{\bibfnamefont{N.}~\bibnamefont{Turok}},
  \bibinfo{journal}{Nucl. Phys.} \textbf{\bibinfo{volume}{B277}},
  \bibinfo{pages}{605} (\bibinfo{year}{1986}).

\bibitem[{\citenamefont{Sousa and Avelino}(2014)}]{Sousa:2014gka}
\bibinfo{author}{\bibfnamefont{L.}~\bibnamefont{Sousa}} \bibnamefont{and}
  \bibinfo{author}{\bibfnamefont{P.~P.} \bibnamefont{Avelino}},
  \bibinfo{journal}{Phys. Rev.} \textbf{\bibinfo{volume}{D89}},
  \bibinfo{pages}{083503} (\bibinfo{year}{2014}), \eprint{1403.2621}.

\bibitem[{\citenamefont{Guedes et~al.}(2018)\citenamefont{Guedes, Avelino, and
  Sousa}}]{Guedes:2018afo}
\bibinfo{author}{\bibfnamefont{G.~S.~F.} \bibnamefont{Guedes}},
  \bibinfo{author}{\bibfnamefont{P.~P.} \bibnamefont{Avelino}},
  \bibnamefont{and} \bibinfo{author}{\bibfnamefont{L.}~\bibnamefont{Sousa}},
  \bibinfo{journal}{Phys. Rev.} \textbf{\bibinfo{volume}{D98}},
  \bibinfo{pages}{123505} (\bibinfo{year}{2018}), \eprint{1809.10802}.

\bibitem[{\citenamefont{Cui et~al.}(2018)\citenamefont{Cui, Lewicki, Morrissey,
  and Wells}}]{Cui:2017ufi}
\bibinfo{author}{\bibfnamefont{Y.}~\bibnamefont{Cui}},
  \bibinfo{author}{\bibfnamefont{M.}~\bibnamefont{Lewicki}},
  \bibinfo{author}{\bibfnamefont{D.~E.} \bibnamefont{Morrissey}},
  \bibnamefont{and} \bibinfo{author}{\bibfnamefont{J.~D.} \bibnamefont{Wells}},
  \bibinfo{journal}{Phys. Rev.} \textbf{\bibinfo{volume}{D97}},
  \bibinfo{pages}{123505} (\bibinfo{year}{2018}), \eprint{1711.03104}.

\bibitem[{\citenamefont{Lorenz et~al.}(2010)\citenamefont{Lorenz, Ringeval, and
  Sakellariadou}}]{Lorenz:2010sm}
\bibinfo{author}{\bibfnamefont{L.}~\bibnamefont{Lorenz}},
  \bibinfo{author}{\bibfnamefont{C.}~\bibnamefont{Ringeval}}, \bibnamefont{and}
  \bibinfo{author}{\bibfnamefont{M.}~\bibnamefont{Sakellariadou}},
  \bibinfo{journal}{JCAP} \textbf{\bibinfo{volume}{1010}}, \bibinfo{pages}{003}
  (\bibinfo{year}{2010}), \eprint{1006.0931}.

\bibitem[{\citenamefont{Blanco-Pillado
  et~al.}(2014)\citenamefont{Blanco-Pillado, Olum, and
  Shlaer}}]{Blanco-Pillado:2013qja}
\bibinfo{author}{\bibfnamefont{J.~J.} \bibnamefont{Blanco-Pillado}},
  \bibinfo{author}{\bibfnamefont{K.~D.} \bibnamefont{Olum}}, \bibnamefont{and}
  \bibinfo{author}{\bibfnamefont{B.}~\bibnamefont{Shlaer}},
  \bibinfo{journal}{Phys. Rev.} \textbf{\bibinfo{volume}{D89}},
  \bibinfo{pages}{023512} (\bibinfo{year}{2014}), \eprint{1309.6637}.

\bibitem[{\citenamefont{Blanco-Pillado and
  Olum}(2019)}]{Blanco-Pillado:2019tbi}
\bibinfo{author}{\bibfnamefont{J.~J.} \bibnamefont{Blanco-Pillado}}
  \bibnamefont{and} \bibinfo{author}{\bibfnamefont{K.~D.} \bibnamefont{Olum}}
  (\bibinfo{year}{2019}), \eprint{1912.10017}.

\bibitem[{\citenamefont{Caldwell and Allen}(1992)}]{Caldwell:1991jj}
\bibinfo{author}{\bibfnamefont{R.~R.} \bibnamefont{Caldwell}} \bibnamefont{and}
  \bibinfo{author}{\bibfnamefont{B.}~\bibnamefont{Allen}},
  \bibinfo{journal}{Phys. Rev.} \textbf{\bibinfo{volume}{D45}},
  \bibinfo{pages}{3447} (\bibinfo{year}{1992}).

\bibitem[{\citenamefont{DePies and Hogan}(2007)}]{DePies:2007bm}
\bibinfo{author}{\bibfnamefont{M.~R.} \bibnamefont{DePies}} \bibnamefont{and}
  \bibinfo{author}{\bibfnamefont{C.~J.} \bibnamefont{Hogan}},
  \bibinfo{journal}{Phys. Rev.} \textbf{\bibinfo{volume}{D75}},
  \bibinfo{pages}{125006} (\bibinfo{year}{2007}), \eprint{astro-ph/0702335}.

\bibitem[{\citenamefont{Blanco-Pillado
  et~al.}(2019)\citenamefont{Blanco-Pillado, Olum, and
  Wachter}}]{Blanco-Pillado:2019vcs}
\bibinfo{author}{\bibfnamefont{J.~J.} \bibnamefont{Blanco-Pillado}},
  \bibinfo{author}{\bibfnamefont{K.~D.} \bibnamefont{Olum}}, \bibnamefont{and}
  \bibinfo{author}{\bibfnamefont{J.~M.} \bibnamefont{Wachter}}
  (\bibinfo{year}{2019}), \eprint{1907.09373}.

\bibitem[{\citenamefont{Avelino and Sousa}(2012)}]{Avelino:2012qy}
\bibinfo{author}{\bibfnamefont{P.~P.} \bibnamefont{Avelino}} \bibnamefont{and}
  \bibinfo{author}{\bibfnamefont{L.}~\bibnamefont{Sousa}},
  \bibinfo{journal}{Phys. Rev.} \textbf{\bibinfo{volume}{D85}},
  \bibinfo{pages}{083525} (\bibinfo{year}{2012}), \eprint{1202.6298}.

\bibitem[{\citenamefont{Blanco-Pillado
  et~al.}(2011)\citenamefont{Blanco-Pillado, Olum, and
  Shlaer}}]{BlancoPillado:2011dq}
\bibinfo{author}{\bibfnamefont{J.~J.} \bibnamefont{Blanco-Pillado}},
  \bibinfo{author}{\bibfnamefont{K.~D.} \bibnamefont{Olum}}, \bibnamefont{and}
  \bibinfo{author}{\bibfnamefont{B.}~\bibnamefont{Shlaer}},
  \bibinfo{journal}{Phys. Rev.} \textbf{\bibinfo{volume}{D83}},
  \bibinfo{pages}{083514} (\bibinfo{year}{2011}), \eprint{1101.5173}.

\bibitem[{\citenamefont{Aghanim et~al.}(2018)}]{Aghanim:2018eyx}
\bibinfo{author}{\bibfnamefont{N.}~\bibnamefont{Aghanim}} \bibnamefont{et~al.}
  (\bibinfo{collaboration}{Planck}) (\bibinfo{year}{2018}),
  \eprint{1807.06209}.

\end{thebibliography}
\end{document}